\documentclass[twocolumn,reprint,amsmath,amssymb,showpacs,prb,aps,superscriptaddress]{revtex4-1}
\usepackage{graphicx}
\usepackage{times}
\usepackage{setspace}
\usepackage{CJK}

\begin{document}
\begin{CJK*}{UTF8}{}

\title{Length and temperature dependent crossover of charge transport across molecular junctions}
\CJKfamily{bsmi}

\author{Ya-Lin Lo (羅雅琳)}
\affiliation{ Department of Physics, National Taiwan University, Taipei 106,Taiwan}
\affiliation{ Center of Theoretical Sciences, National Taiwan University, Taipei 106, Taiwan}

\author{ Shih-Jye Sun (孫士傑)}
\email{sjs@nuk.edu.tw}
\affiliation{Department of Applied Physics, National University of Kaohsiung, Kaohsiung 811, Taiwan}

\author{Ying-Jer Kao (高英哲)}
\email{yjkao@phys.ntu.edu.tw}
\affiliation{ Department of Physics, National Taiwan University, Taipei 106,Taiwan}
\affiliation{ Center of Theoretical Sciences, National Taiwan University, Taipei 106, Taiwan}
\affiliation{ Center for Quantum Science and Engineering, National Taiwan University, Taipei 106, Taiwan}

\date{\today}
\begin{abstract}
 We study the electronic transport in a molecular junction, in which each unit is coupled to a local phonon bath,  using the non-equilibrium Green's function method. We observe the conductance oscillates with the molecular chain length and the oscillation period in odd-numbered chains depends strongly on the applied bias. This oscillatory behavior is smeared out at the bias voltage near the phonon energy. For the phonon-free case, we find a crossover from tunneling to thermally activated  transport as the length of the molecule increases.  In the presence of electron-phonon interaction, the transport is  thermally driven and a crossover from the thermally suppressed to assisted conduction is observed. 

 \end{abstract}
\pacs{72.60.+g,	
73.23.-b,	
73.63.-b,	
81.07.Nb	
}

\maketitle
\end{CJK*}

\newcommand{\figuremacroW}[3]{
	\begin{figure}[tbhp]
		\centerline{\includegraphics[width=3in]{#1}}
		\caption{#2}
		\label{#3}
	\end{figure}
}
\section{INTRODUCTION}

The research on the molecular electronics is  a fast progressing field due to its great application potentials. Molecular junctions  with  single molecules or molecular assemblies sandwiched between two electrodes are of particular interests since they are potential candidates as the building blocks for the next generation of electronics.\cite{tao_review} One important issue associated with the electron transport in molecular devices is understanding the role  phonons play in the conduction mechanism. In general, the electron-phonon interaction is expected to  modify transport properties when the phonon relaxation time is comparable to the typical transport time of the junction.\cite{nitzen_review_2003_science}  
Thanks to the rapid development of new experimental techniques, it is now possible to study quantum transport phenomena in a single-molecule wire experimentally using break junctions,\cite{1997_breakjunction} STM,\cite{2001_stm_monolayer,1995_stm_coulomb,2000_stm_nature} and crossed-wire tunnel junctions.\cite{2002_crossed_wire}  Vibronic contributions to the charge transport have been identified in the low-bias conductivity of molecular wires,\cite{2004_nano_lett_iets,2004_nano_lett_iets2} step-like features in current-voltage characteristics\cite{Nanotechnology_2003_gated_devices,steplike_metal,PhysRevLett.92.206102,PhysRevB.77.113405,PhysRevLett.88.226801} and temperature-dependent conduction.\cite{termal_active_2004,termal_effects,PhysRevLett.86.288} 
It has been argued that  the transmission rate of the electron decays exponentially with length for short chains, which is a signature of  tunneling transport.\cite{REVIEW1}  For long chains, however, thermal excitations dominate at high temperature and  a crossover to thermally activated diffusion with Ohmic behavior best describes the electron transfer in this regime.\cite{noph1,noph2}
 In molecular junctions with fixed molecular length, a crossover from coherent to incoherent conduction, as temperature increases, has been observed;\cite{PhysRevB.73.075428,PhysRevB.66.075303,PhysRevB.71.235116} in addition,  strong experimental evidences demonstrated that there is a crossover  from tunneling to hopping  of the charge transport mechanism as the length of the molecule increases. \cite{tuneling_to_hopping_2008_scinece_OPI,Lu_2009_OPE,Frisbie_tTOh_2010_JACS_OPT} It is not clear, however, what is the underlying mechanism of this crossover and how it is modified in the presence of electron-phonon interactions.

 In this paper, we study electrons interacting with local phonon baths in the molecular wires and we treat the on-site electron-phonon interaction perturbatively to the second order in the coupling strength. This may become important in molecules which couples to vibration modes provided by the local functional group and the charge transport characteristics can change significantly.   
 Using the standard non-equilibrium Green's function (NEGF) method and a tight-binding model for the molecular chain,  we study the  conduction mechanism as a function of the length and the temperature.  
We observe  a crossover of the electron conduction mechanism  from tunneling to thermally activated transport as the molecular length increases in the absence of the electron-phonon interaction. In the presence of the electron-phonon coupling, a crossover from thermally suppressed  to assisted conduction  is also observed. 

The paper is organized as follows. In Sec.~\ref{method}, we describe  the microscopic model  of the molecular wire, and the
theoretical framework to compute the transport properties in the presence of the electron-phonon interaction. In Sec.~\ref{result}, we show
our results and interpret their physics based on our proposed model and approximations. Finally, we conclude in Sec.~\ref{conclusion}.

\section {MODEL AND THEORETICAL METHOD} \label{method}
\figuremacroW{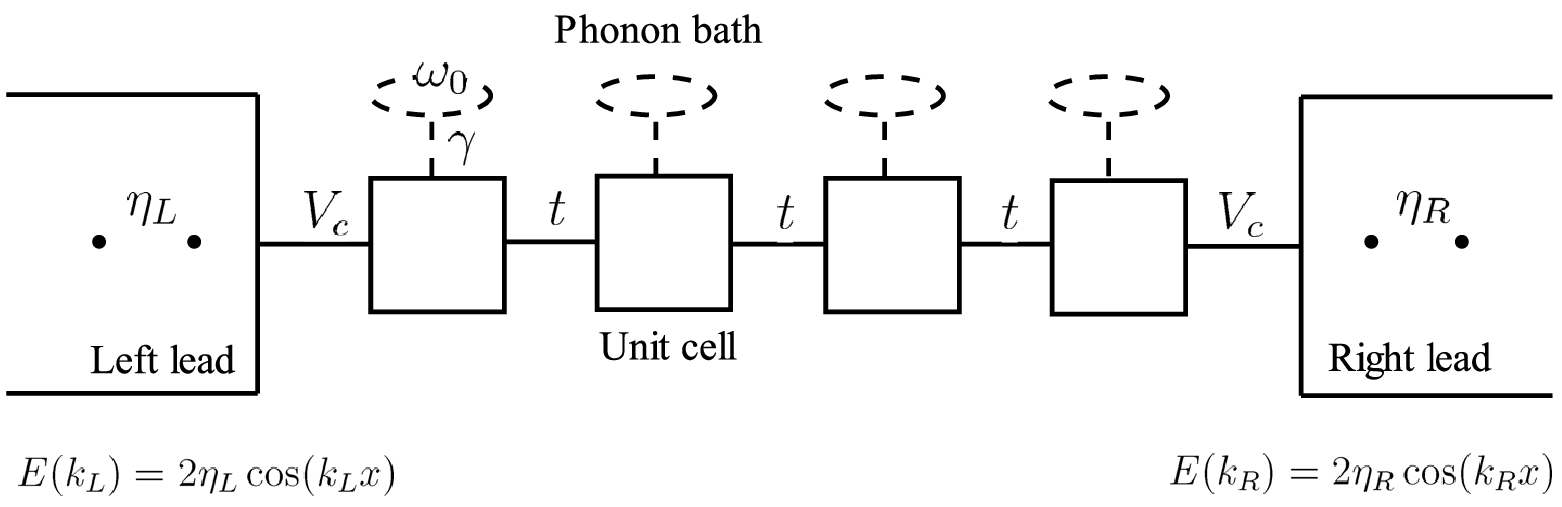}{One molecular chain, sandwiched in two leads, consists of $N$ units and each unit is coupled to a local phonon bath.}{setup}
We model the molecule as a one dimensional chain of $N$ units and each unit is coupled locally to a phonon bath (Fig.~\ref{setup}). 
The full Hamiltonian of the molecular system with the local electron-phonon interaction can be written as
\begin{equation}\label{Ham}
    H=H_l+H_c+H_{el}+H_{ph}+H_{e-ph}\;.
\end{equation}
We assume the two semi-infinite one dimensional leads, and 
 the lead Hamiltonian is given by,
\begin{equation}\label{HL}
    H_l=\sum_{\alpha\in {L,R} }\sum_{i=1}^{\infty}  \epsilon_{\alpha} c^{\dagger}_{i,\alpha}c_{i,\alpha}+\eta_\alpha\sum_{\langle i,j\rangle}c^{\dagger}_{i,\alpha}c_{j,\alpha}\;,
\end{equation} 
where $c^\dagger_{i,\alpha}$ ($c_{i,\alpha}$) is creation (annihilation) operator of the carrier at the $i$th unit in the  lead $\alpha=R, L$ , $\eta_\alpha$ is the hopping integral in the leads, and $\epsilon_{\alpha}$ is the on-site energy in the $\alpha$ lead.  The coupling between leads and the molecule is given by,
 \begin{equation}\label{HC}
    H_c=\sum_{\alpha \in {L,R},i \in {1,N}} V_{\alpha}(c^{\dagger}_{1,\alpha}d_{i}^{\phantom{\dagger}}+h.c)\;,
\end{equation}
where the $L(R)$ lead only couples to the first ($N$th) unit of the molecule via the
coupling constant, $V_{L(R)}$ , and $d^\dagger_i$ ($d_i$) is creation  (annihilation) operator of  the carrier at the $i$th unit of the middle wire. The Hamiltonian for the molecule is given by a single-orbital tight-binding model, 
\begin{equation}\label{He}
    H_{el}=\sum_{i=1}^N  \epsilon_i d^{\dagger}_id_{i}+t\sum_{\langle i,j\rangle} d^{\dagger}_{i}d_j^{\phantom{\dagger}} \;,
\end{equation}
where $t$ is the hopping integral between the nearest-neighbor units, $\epsilon_i$ is the orbital energy
at the $i$th molecular unit. 

For simplicity, we set  the vibrational mode  $\omega=\omega_0$ of the phonon baths to be the same  and the electron-phonon interaction is given by,
 \begin{equation}\label{Hp}
    H_{ph}=\sum_{i=1}^N \omega_0 b^{\dagger}_i b_i\;,
\end{equation}
where $i$ is unit index and $b_i$ ($b^{\dagger}$) is the phonon annihilation (creation) operator, and 
\begin{equation}\label{Hep}
    H_{e-ph}=\gamma\sum_{i=1}^N d^{\dagger}_i d_{i}(b_i^{\dagger}+b_i)\;,
\end{equation}
where $\gamma$ is the strength of  the electron-phonon coupling.

The current through an interacting region driven by a bias voltage is obtained by the Keldysh NEGF, \cite{general_current}
\begin{widetext}
\begin{equation}
 I=\frac{ie}{2h}\int d\omega \;{\rm Tr}\{  
  [f_{L}(\omega,\mu_L)\mathbf{\Gamma}_{L}-f_{R}(\omega,\mu_R)\mathbf{\Gamma}_{R}]
   [\mathbf{G}^{r}(\omega)-\mathbf{G}^{a}(\omega)]
  +[\mathbf{\Gamma}_{L}-\mathbf{\Gamma}_{R}] \mathbf{G}^{<}(\omega)\}\;,
  \label{eq:current}
\end{equation}
\end{widetext}
where $\mathbf{\Gamma}_{\alpha\in L,R}$ are related to the imaginary parts of the self-energies of leads, $\mathbf{\Gamma}_{\alpha\in L,R}= (\mathbf{\Sigma}^a_\alpha(\omega)-\mathbf{\Sigma}^r_\alpha(\omega))/2 i$,\cite{haug} and 
$f_\alpha(\omega,\mu_\alpha)=1/(1+e^{\beta (\omega-\mu_\alpha)})$ are Fermi-Dirac distributions in  the reservoirs.

 The full (retarded, advanced and lesser) 
 Green's functions are obtained by solving the Dyson equation,
\begin{eqnarray}
\mathbf{G}=\mathbf{g}+\mathbf{g}\mathbf{V}\mathbf{G}\;,
\end{eqnarray}
where $\mathbf{G}$ is the full Green's function  with the electron-phonon interaction, $\mathbf{g}$ is the non-interacting Green's function, and $\mathbf{V}$ is the interaction part. Following the notations in Ref.~\onlinecite{woh},
we solve the Dyson equation by keeping terms up to the second order in the electron-phonon coupling $\gamma$ to obtain the full Green's function, 
\begin{eqnarray}
\mathbf{G}=\mathbf{g}(\mathbf{I}+\gamma \mathbf{G}^Q)\;,
\end{eqnarray}
where $\mathbf{G}^Q$ contains all the information about the electron-phonon interaction,
\begin{eqnarray}
G^Q_{ij}= \left\langle \left\langle d_i(b^{\phantom{\dagger}}_i+b_i^\dagger);d_j^\dagger\right\rangle\right\rangle \delta_{i,j} \;,
\end{eqnarray}
where we use superscript $Q$ to indicate the presence of the phonon operators $b$ and $b^\dagger$.  $\langle\langle a;b^\dagger\rangle\rangle$ represents  one of the Keldysh Green's functions with (composite) operators $a$ and $b^\dagger$, and their Fourier transforms are defined as 
\begin{eqnarray}
G^<_{ab}(\omega)&=&i\int dt e^{i\omega t}\langle b^\dagger a(t)\rangle,\nonumber\\
G^>_{ab}(\omega)&=&-i\int dt e^{i\omega t}\langle a(t) b^\dagger\rangle,
\end{eqnarray}
for the lesser and greater Green's functions, and
\begin{eqnarray}
G^r_{ab}(\omega)&=&-i\int_0^\infty dt e^{i(\omega+i0^+) t}\langle [a(t), b^\dagger]_+\rangle,\nonumber\\
G^a_{ab}(\omega)&=&i\int_{-\infty}^0 dt e^{i(\omega-i0^+) t}\langle [a(t), b^\dagger]_+\rangle,
\end{eqnarray}
for the retarded and advanced Green's functions.
Moreover, in our model we consider only the on-site electron-phonon interaction; therefore,
 $\mathbf{G}^Q$  is diagonal, and can be written as
 \begin{eqnarray}
G^{Q}_{ij}=g_{ij}\left(\left\langle b_i+b_i^\dagger \right\rangle+\gamma G^{QQ}_{ii}\right)\delta_{ij}\;,
\end{eqnarray}
where 
\begin{eqnarray}
\label{giqi}
G^{QQ}_{ii}= \left\langle \left\langle d_i(b_i+b_i^\dagger);(b_i+b_i^\dagger)d_i^\dagger\right\rangle\right\rangle \;.
\end{eqnarray}
The retarded Green's function, $\mathbf{G}^{QQ,r}$, can be evaluated as,
\begin{widetext}
\begin{align}
\nonumber
G^{QQ,r}_{ii}(\omega) &=-i\int^\infty_{0}dt e^{i\omega t}\langle[ d_i(b_i+b^\dagger_i);(b_i+b_i^\dagger)d_i^\dagger]_+\rangle_t \\
&= i\int \frac{d\omega^{'}}{2\pi}g^{>}_{i,i}(\omega^{'})
  \left(\frac{\langle b_ib_i^\dagger\rangle}{\omega-\omega_0-\omega^{'}+i\delta}+\frac{\langle b_i^\dagger b_i\rangle}{\omega+\omega_0-\omega^{'}+i\delta }\right)-g^{<}_{i,i}(\omega^{'})\left(\frac{\langle b_ib_i^\dagger\rangle}{\omega+\omega_0-\omega^{'}+i\delta}+\frac{\langle b_i^\dagger b_i\rangle}{\omega-\omega_0-\omega^{'}+i\delta
   }\right) \;.
\label{eq:G_QQ}
\end{align}
\end{widetext}
Collecting all the terms and  the inverse of the full Green's function is given as
\begin{align}
\mathbf{G}^{-1}&=\mathbf{g}^{-1}-\mathbf{\Sigma}_{ph}\;,\\
\mathbf{\Sigma}_{ph}&=\gamma\left\langle b_i^{\phantom{\dagger}}+b_i^\dagger \right\rangle \delta_{i,j}+\gamma^2 \mathbf{G}_{QQ}=\mathbf{\Sigma}^1_{ph}+\mathbf{\Sigma}^2_{ph}\;,
\label{eq17}
\end{align}
where the bare Green's function $\mathbf{g}$ is defined as the non-interacting electron Green's function together with the self-energy corrections from the leads,
\begin{eqnarray}
\mathbf{g}=(\omega-\mathbf{H}_{el}-\mathbf{\Sigma}_{lead})^{-1}\;.
\end{eqnarray}
The lead self-energy, $\mathbf{\Sigma}_{lead}$ , describes the interaction between the leads and the molecular chain, and  for  semi-infinite one-dimensional leads,\cite{mujica:6849,PhysRevB.72.115439}
\begin{eqnarray}
\mathbf{\Sigma}_{lead}&=&\sum_{\alpha\in R,L} \mathbf{\Sigma_{\alpha}} \;,\\
\label{leadselfenergies}
\mathbf{\Sigma}_{\alpha}^{r,a}&=& 
\frac{|V_{\alpha}|^2}{ \frac{\omega-\epsilon_{\alpha}}{2}\mp i\sqrt{\eta_\alpha^2-(\frac{\omega-\epsilon_{\alpha}}{2})^2}} \;\delta_{i,j}\;\delta_{i,n}\;,\\
\mathbf{\Sigma}_{\alpha}^{<}&=& i\mathbf{\Gamma}_{\alpha} f_{\alpha}(\omega)\;,  \label{lesser_self_lead} \\
\mathbf{\Sigma}_{\alpha}^{>}&=& -i\mathbf{\Gamma}_{\alpha} [1-f_{\alpha}(\omega)] \;,
\end{eqnarray}
where $n$ is  $1$ ($N$) for $\alpha=L(R)$.  
One remaining quantity required to evaluate the current is the full lesser Green's function, $\mathbf{G}^{<}$ , and it can be calculated through the $\mathbf{G}^r$ and $\mathbf{G}^a$,
\begin{eqnarray}
\mathbf{G}^{<}= \mathbf{G}^{r}\mathbf{\Sigma}^{<}\mathbf{G}^{a}\;,\\
\mathbf{\Sigma}^{<}=\mathbf{\Sigma}^{<}_{lead}+\mathbf{\Sigma}^{<}_{ph}\;,
\end{eqnarray}
where the full lesser self-energy is broken up into contributions from the leads and the phonon, and the former is given by Eq.~(\ref{lesser_self_lead}).
For the phonon self-energy, 
$\mathbf{\Sigma}^1_{ph}$ in Eq.~(\ref{eq17}), gives a shift in the energy. To evaluate the expectation value of the phonon operators, we perturb the phonon wavefunction to the first order in $\gamma$,
\begin{eqnarray}\label{wfun}
\nonumber  
|\Psi_{n^{'}_i}\rangle &=& |n^{'}_i\rangle+\sum_{n\neq n^{'}}\frac{\langle n_i |H_{e-ph}|n^{'}_i\rangle}{E_n^{'}-E_n}|n_i\rangle\;, \\
   &=& |n^{'}_i\rangle +\frac{\gamma}{ \omega_0} d^{+}_i d_i  \sum_{n\neq n^{'}}\frac{\langle
   n_i|b_i+b_i^{+}|n^{'}_i\rangle}{n^{'}-n}|n_i\rangle\;,
\end{eqnarray}
where $|n^{'}_i\rangle$ and $|n_i\rangle$ are phonon states and the energy difference is proportional to the difference of their phonon quantum numbers. 
\begin{eqnarray}
\left\langle b_i^{\phantom{\dagger}}+b_i^\dagger \right\rangle =i\frac{2\gamma}{\omega_0}\int\frac{d\omega}{2\pi}\mathbf{g}^{<}\;.
\end{eqnarray}
The second term $\mathbf{\Sigma}^{<,2}_{ph}=\gamma^2 \mathbf{G}^{QQ,<}$, where $\mathbf{G}^{QQ,<}$ is diagonal with matrix elements,
\begin{align}
\nonumber
 G_{ii}^{QQ,<}(\omega) &=i\int dt e^{i\omega t}\langle(b_i+b_i^\dagger) (b_i(t)+b^\dagger_i(t))\rangle \langle d_i^\dagger d_i(t)\rangle\;,\\
 &=\left\langle b_i^\dagger b_i\right\rangle g_{ii}^{<}(\omega-\omega_0)+\left\langle b_i b_i^\dagger\right\rangle g_{ii}^{<}(\omega+\omega_0)\;,
\end{align}
where $\langle b^\dagger b\rangle = (e^{\beta \omega_0}-1)^{-1}$ is  the phonon occupation number.

\figuremacroW{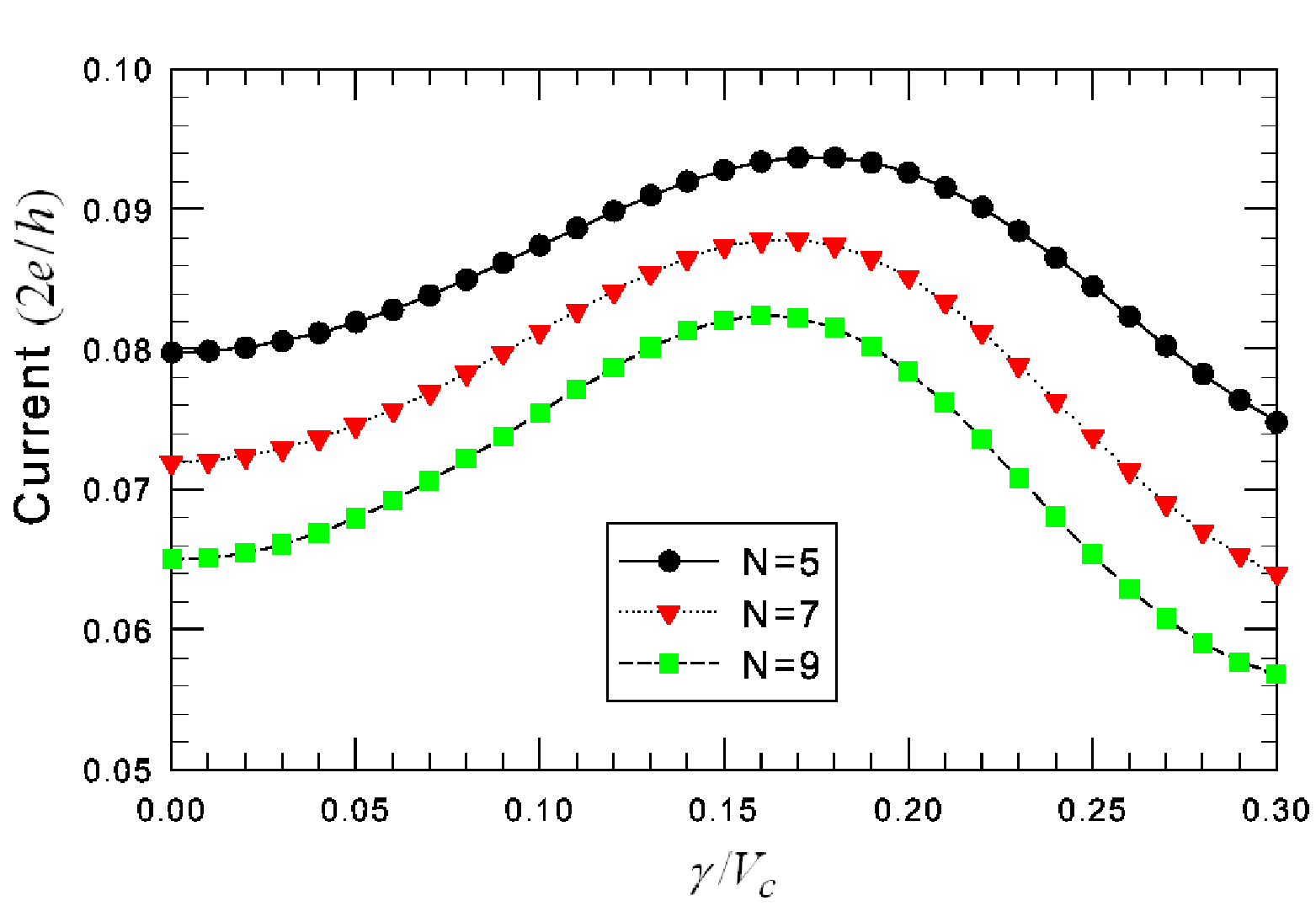}{(Color online) Current as a function of $\gamma$ at $\omega_0=0.4$ for different odd-numbered chains: $N=5$ (circle), 7 (triangle), and 9 (square)  at $V_a=0.1$. The injection electron hits a resonance state at the corresponding $\gamma$ to produce the maximum.}{iga_va01_04hw_odd}  

\section{RESULTS and DISCUSSION} \label{result}
 We consider the simplest case of the model, Eq.~(\ref{HL}), where  two electrodes are  the identical, $\eta_{L}=\eta_{R}\equiv \eta$ , and have the same coupling to the molecule, $V_{L}=V_{R}\equiv V_c=1$. In the following, unless otherwise noticed, the energy will be in units of $V_c$. 
The parameters used in this paper are $\epsilon_{L,R}=\epsilon_i=0,\eta=-5,t=-2 $. The symbol $T$ indicates the temperature both in the reservoirs and phonon baths. The conductance is in units of $G_0=2e^2/h$, the current is in units of $I_0=2e/h$ , and the resistance is in units of  $R_0=1/G_0$. 
The current in Eq.~(\ref{eq:current}) is driven by a bias voltage $ V_a$. We assume that the left lead is connected to the ground, with the chemical potentials $\mu_L=0, \mu_R=-e\,V_a$.  We obtain the Green's functions, for example, in Eq.~(\ref{eq:G_QQ}) by performing numerical integration in the real frequency domain, and  the small positive $\delta$ is set to $0.01$. 

\subsection{Even-odd effect}
We start by considering  short chains first.  Figures~\ref{iga_va01_04hw_odd} and \ref{iga_va01_04hw_even} show the behavior of currents at small $\gamma$ for chains with odd and even sites at $T=0.025$, and the phonon mode is at $\omega_0=0.4$. This case is in the quantum limit as $T/\omega_0\ll 1$. First we observe that odd-numbered chains shows significant larger current than even-numbered chains, as found in Ref.~\onlinecite{oscillatory} and Ref.~\onlinecite{PhysRevB.60.6028}.  For even-numbered chains, in the absence of electron-phonon interaction ($\gamma=0$) and at small bias, the chemical potential of the leads falls in the gap between the highest occupied molecular orbital (HOMO) and the lowest unoccupied molecular orbital (LUMO), thus the tunneling current is small. On the other hand, for odd-numbered chains, there exists a state close to the lead chemical potential, and resonant tunneling is possible, which produces a finite current.\cite{oscillatory,PhysRevB.60.6028} 
The electron-phonon interaction has a strong effect on the conductance of the odd-numbered chains.  The molecular state near the lead chemical potential is shifted by the interaction, and a resonance occurs at a finite $\gamma$ (Fig.~\ref{iga_va01_04hw_odd}). The resonance is not observed in the even-numbered chains, and the conductance increases monotonically as $\gamma$ increases.
One expects this even-odd effect should vanish when the number of sites becomes large as the HOMO-LUMO gap decreases with more atoms in the molecular wire. 
Moreover, this transport behavior  depends strongly on these physical parameters, the Fermi level of the leads and the HOMO and LUMO of the molecules. Thus, one expects this even-odd effect would be tuned  as these physical parameters are changed, and the effects are reversed.

\figuremacroW{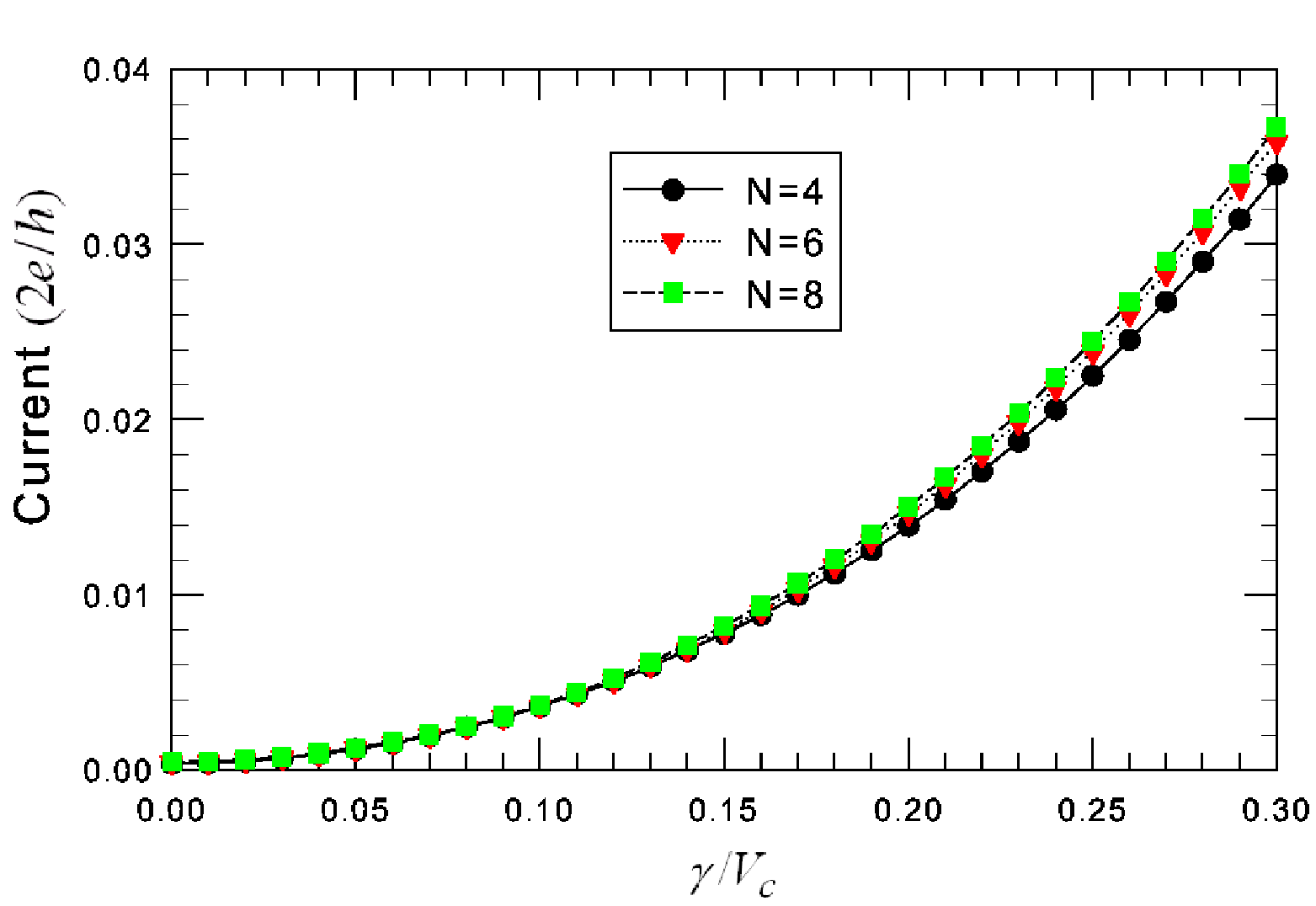}{(Color online) Current as a function of the electron-phonon interaction with $\omega_0= 0.4$ for different even-numbered chains: $N=4$ (circle), $6$ (triangle), and $8$ (square) at $V_a=0.1$.}{iga_va01_04hw_even}   
\figuremacroW{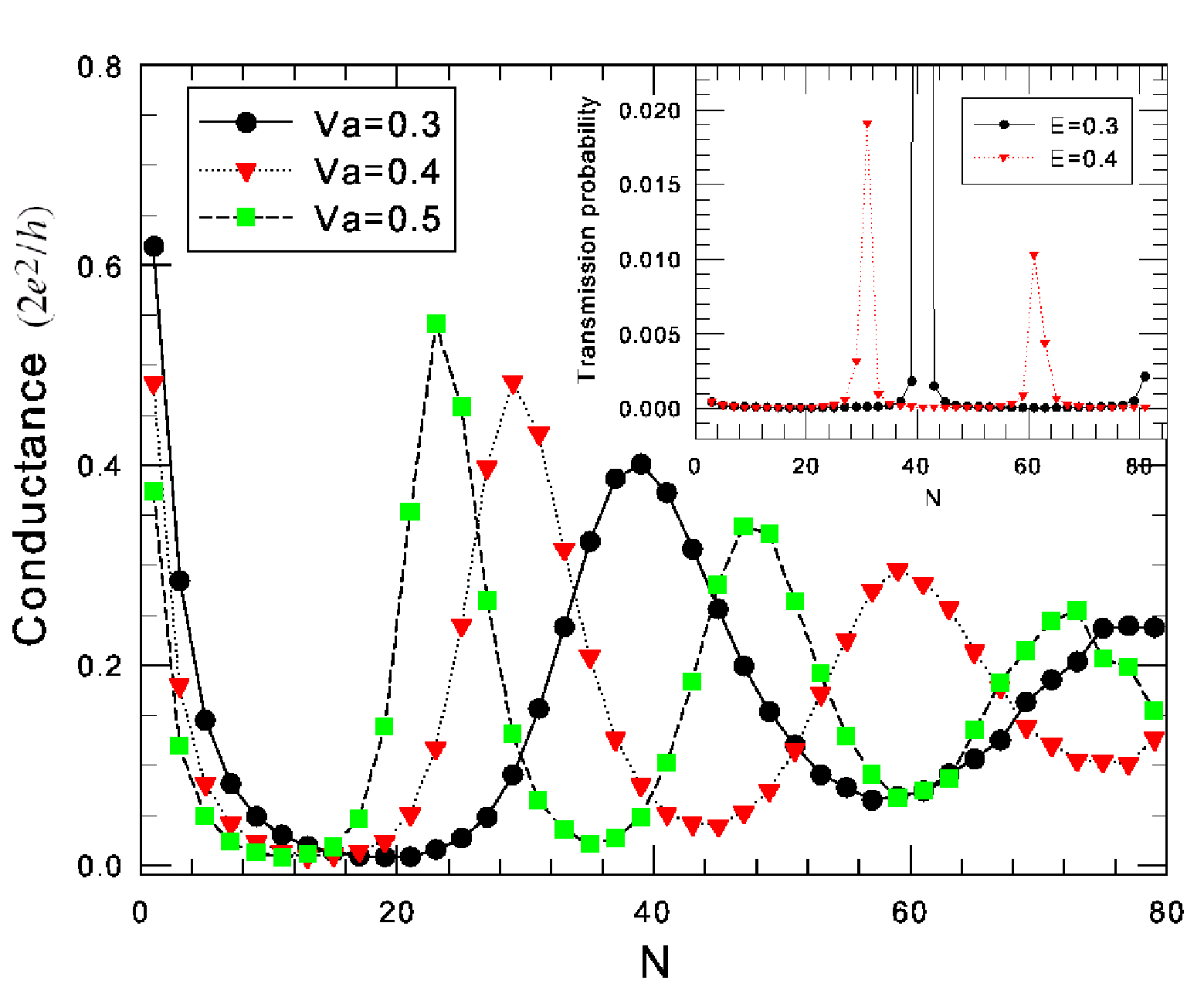}{(Color online) Conductance as a function of length with $\gamma=0$ for different bias voltages $V_a=0.3,\,0.4,\,0.5$.  
Inset shows the maxima of the transmission probability are at $N=31,\,61$ and $N=41$ for $E=0.4$ and $E=0.3$ , respectively.}{ga0-gL-04hw-odd}   

\subsection{Length dependence}
 We will focus on odd-numbered chains  in the following discussion. 
 Figure~\ref{ga0-gL-04hw-odd} shows the conductance versus  the chain length at bias voltages $V_a=0.3,\, 0.4,\,0.5$. It is clear that the conductance also shows oscillatory behavior as the chain length increases, consistent with  literature results.\cite{PhysRevB.72.085431,exposcilation} The period of the oscillation is roughly inverse proportional to the  injection energy of the lead electron. This should be distinguished from the oscillatory behavior in the even-odd effect as discussed in the previous section since this behavior can not be easily explained by the simple argument whether a state is present near the Fermi energy.  The  conductance  maxima   and minima can be  attributed to  the constructive and destructive interferences of the outgoing and reflected wave functions, which  can be better understood by solving the coupled Schr\"{o}dinger equations\cite{Trugman,Emberly} for transmission probabilities as a function of the electron injection energy. Details of the calculation are given in the Appendix. Inset of Fig.~\ref{ga0-gL-04hw-odd} shows the transmission probability as a function of the chain length with external energy $E=0.3,\,0.4$, and the resonances occur at roughly the same molecular length as in the Green's function calculation, confirming the above statement. 

\begin{figure*}[tbhp]
\centerline{\includegraphics[width=7in]{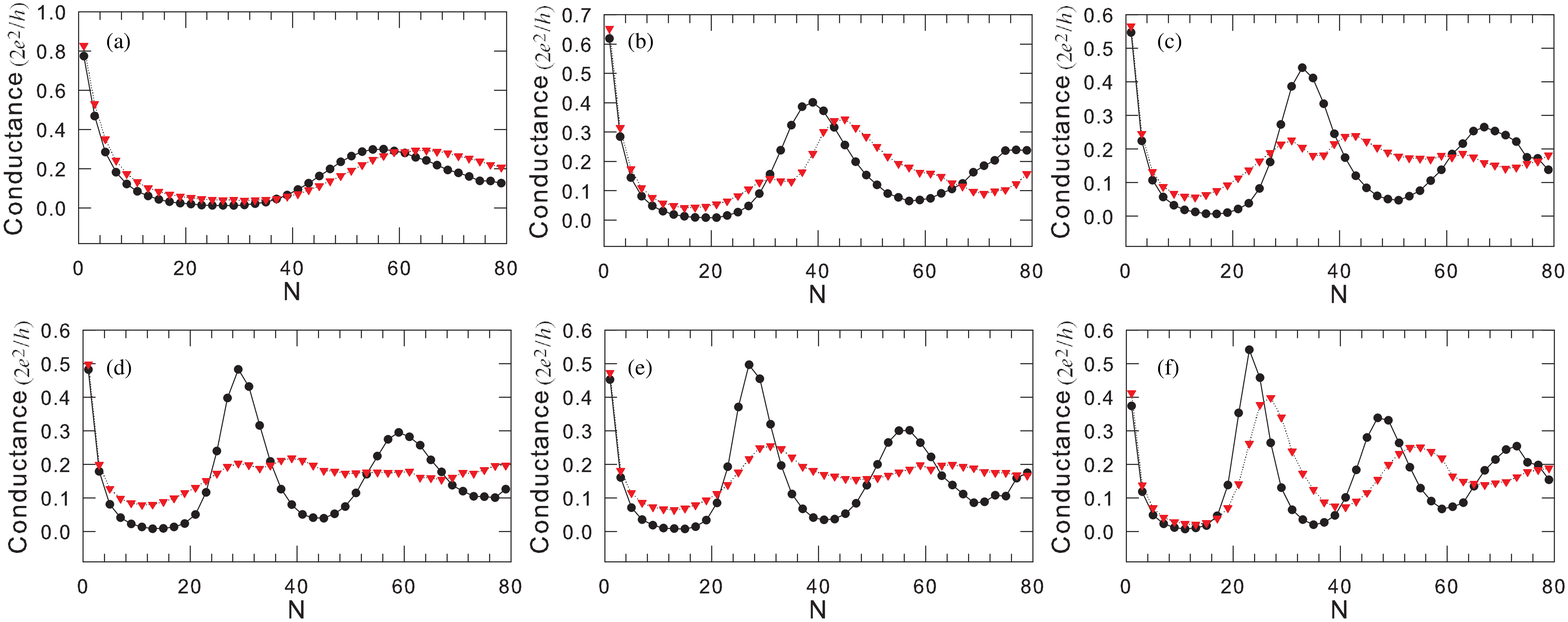}}
	\caption{(Color online) Conductance as a function of length at $V_a=$(a) 0.21,(b) 0.31, (c) 0.36, (d) 0.41, (e) 0.44, and (f) 0.51 for $\gamma=0$ (circles) and $\gamma=0.1$ (triangles) with $\omega_0=0.4$.}
	\label{gL_04hw_va03_odd}
\end{figure*}
\figuremacroW{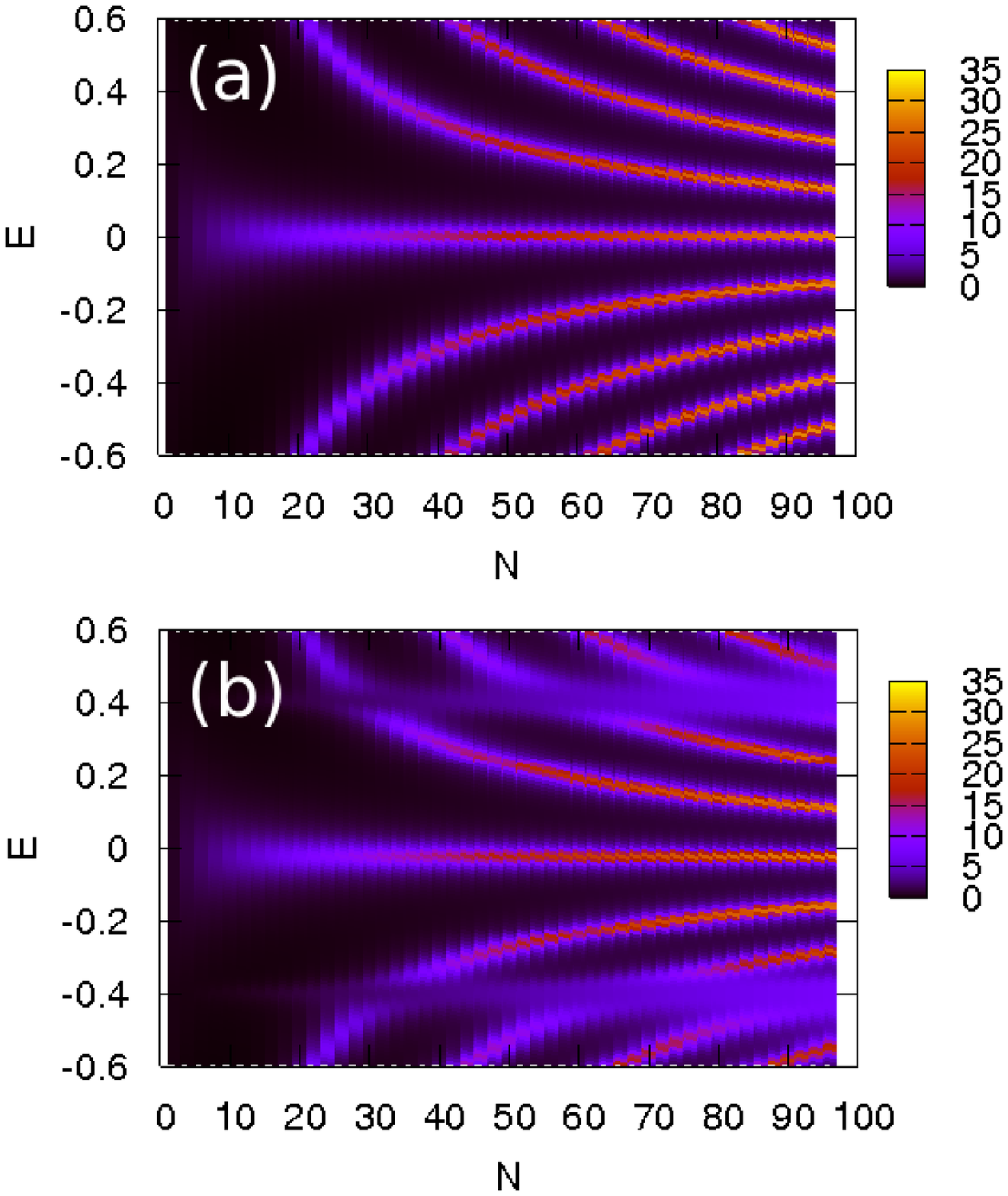}{(Color online) Density of states for molecular with $N$ sites in (a) the phonon-free and (b) $\gamma=0.1,\omega_0=0.4$ cases.}{fig6}
Figure~\ref{gL_04hw_va03_odd} shows the conductance as a function of length with the phonon frequency $\omega_0=0.4$, and electron-phonon coupling $\gamma=0.1$ at various of biases. At low bias, the electron-phonon interaction has little effects on the length dependence. As the bias increases ($V_a$=0.31),  the maximum is suppressed  and the peak position starts to shift. When the external bias is close to the phonon frequency ($V_a$=0.36, 0.41), signature of the oscillation is completely washed out. Further increasing the bias above the phonon frequency, the oscillatory behavior reappears. 
Figure~\ref{fig6} shows the total density of states for the molecules with different length $N$, 
\begin{equation}
D(E)=-{\rm Tr} ({\rm Im} \mathbf{G}^r(E)),
\end{equation}
which roughly corresponds to the conductance  at bias $V_a=E$ (see Eq.~(\ref{eq:current})). Comparing Fig.~\ref{fig6} (a) and (b), it is clear that near $E=\omega_0=0.4$ the density of states is significantly modified by the electron-phonon interaction.  Far away from this energy, the density of states between the two cases are almost identical, and the oscillatory behavior reappears.  
Since we in the quantum limit where $T/\omega_o\ll 1$,  there are  few thermally excited phonons present at low bias. Increasing the bias to larger than $\omega_0$, physical phonons can be excited and inelastic transport process becomes available.  However, these physically excited phonons act as impurities by adding an onsite potential,\cite{Gogolin} and the constructive and destructive interference patterns reappear. On the other hand, close to the threshold of the phonon frequency, one expects large fluctuations of the phonon number, and the phase coherence is lost and the interference pattern disappears.

\subsection{Crossover of the conduction mechanism}
It has been observed in experiments that the conduction mechanism of conjugated molecular wires  changes from tunneling to hopping as the length of the molecule increases in  oligophenyleneimine (OPI),\cite{tuneling_to_hopping_2008_scinece_OPI}  oligonaphthalenefluoreneimine (ONI),\cite{Frisbie_tTOh_2010_JACS_ONI}   oligo($p$-phenylene ethynylene)s (OPE),\cite{Lu_2009_OPE} and oligophenylenetriazole (OPT)\cite{Frisbie_tTOh_2010_JACS_OPT}.
At low bias voltage,  tunneling transport dominates for short chains and the conductance  $G$ is described by Magoga's law,\cite{magoga_1997} 
\begin{align}
G=G_0 e^{-\beta N}, \label{magogalaw}
\end{align}
where $G_0=2e^2/h$,   $N$ is the molecular length, and $\beta$ is the material-dependent inverse decay length. Experimentally, one way to distinguish the tunneling transport from the thermally activated transport  is to compare the $\beta$ values at different regimes.\cite{review_Galperin1} 
For example, in the conjugated OPI molecule,\cite{tuneling_to_hopping_2008_scinece_OPI}   a larger $\beta=3.0$nm$^{-1}$ is observed in the short wire, while in the long chain, a comparatively smaller value of  $\beta=0.9$nm$^{-1}$ is observed.   It is argued that two different  $\beta$ values for   molecular wires of different lengths based on the same oligomers  imply that there is a change in the conduction mechanism. In contrast, in the thermally activated  regime, the resistance increases only linearly with the length  and  the conductance has a strong temperature dependence.
Therefore, we use the following criteria to distinguish tunneling and thermally activated transport in  molecular wires:\cite{hoping.activation.transport}
(1) Large $\beta$ with weak temperature dependence of the conductance corresponds to  tunneling, and small $\beta$ with strong temperature dependence corresponds to thermally activated transport.
(2) Linear length  dependence of the \textit{ resistance} corresponds to thermally activated transport, and exponential decay of the \textit{conductance} corresponds to tunneling.

 Figure~\ref{tunneling_Conductance_ga01_hw02_va01_2units} shows the conductance versus length with fits of $\beta$'s in different regimes.  First we discuss the case where the electron-phonon interaction is absent. Tunneling transport dominates in short chains with a larger $\beta= 0.11$ (per unit), while thermally activated transport  prevails in long chains ($N\geq 21$)  with a smaller $\beta=0.04$ and a linear growth of resistance with length (Fig.~\ref{hopping_Resistance_ga01_hw02_va01_2units}). In addition, in  the long-chain molecule such as $N=31$ (Fig.~\ref{temperature_dependence_2units} (a)),  a crossover from tunneling to thermally activated transport is observed as the temperature increases. At high temperatures, conductance grows as the temperature increases, which is a characteristic of the thermally activated transport conduction. The conductance is less sensitive to the temperature at low temperature, indicating the dominant transport is due to tunneling. For a  short chain $N=5$, in contrast,  the conductance is temperature independent in the whole temperature regime, suggesting  tunneling conduction.
 \figuremacroW{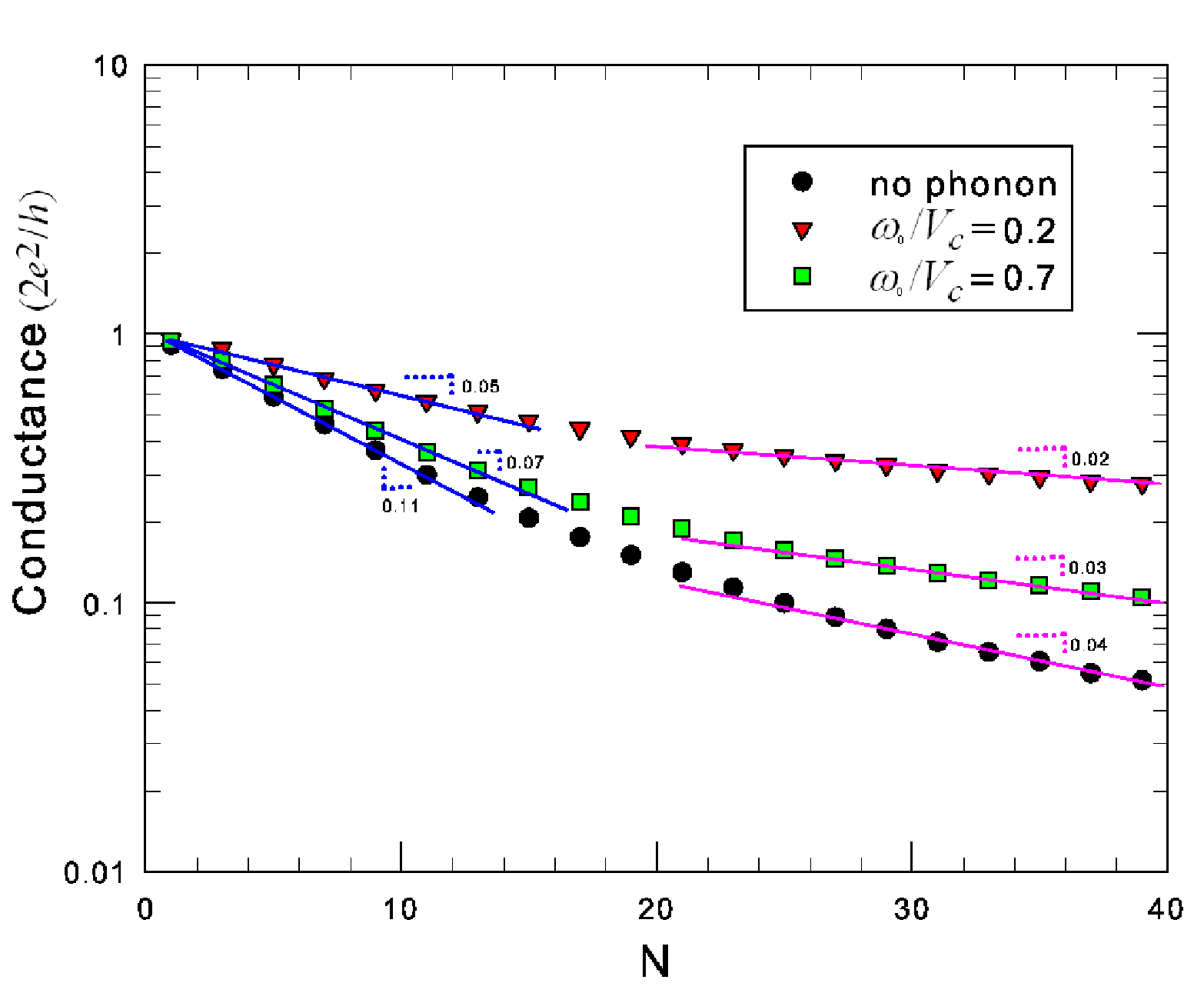}{(Color online) Conductance vs length at $V_a=0.1$. Triangles (squares) are for $\gamma=0.1$, and $\omega_0= 0.2$ (0.7). Circles are for $\gamma=0$. The slope corresponds to the $\beta$ in Eq.~(\ref{magogalaw}).}{tunneling_Conductance_ga01_hw02_va01_2units}
\figuremacroW{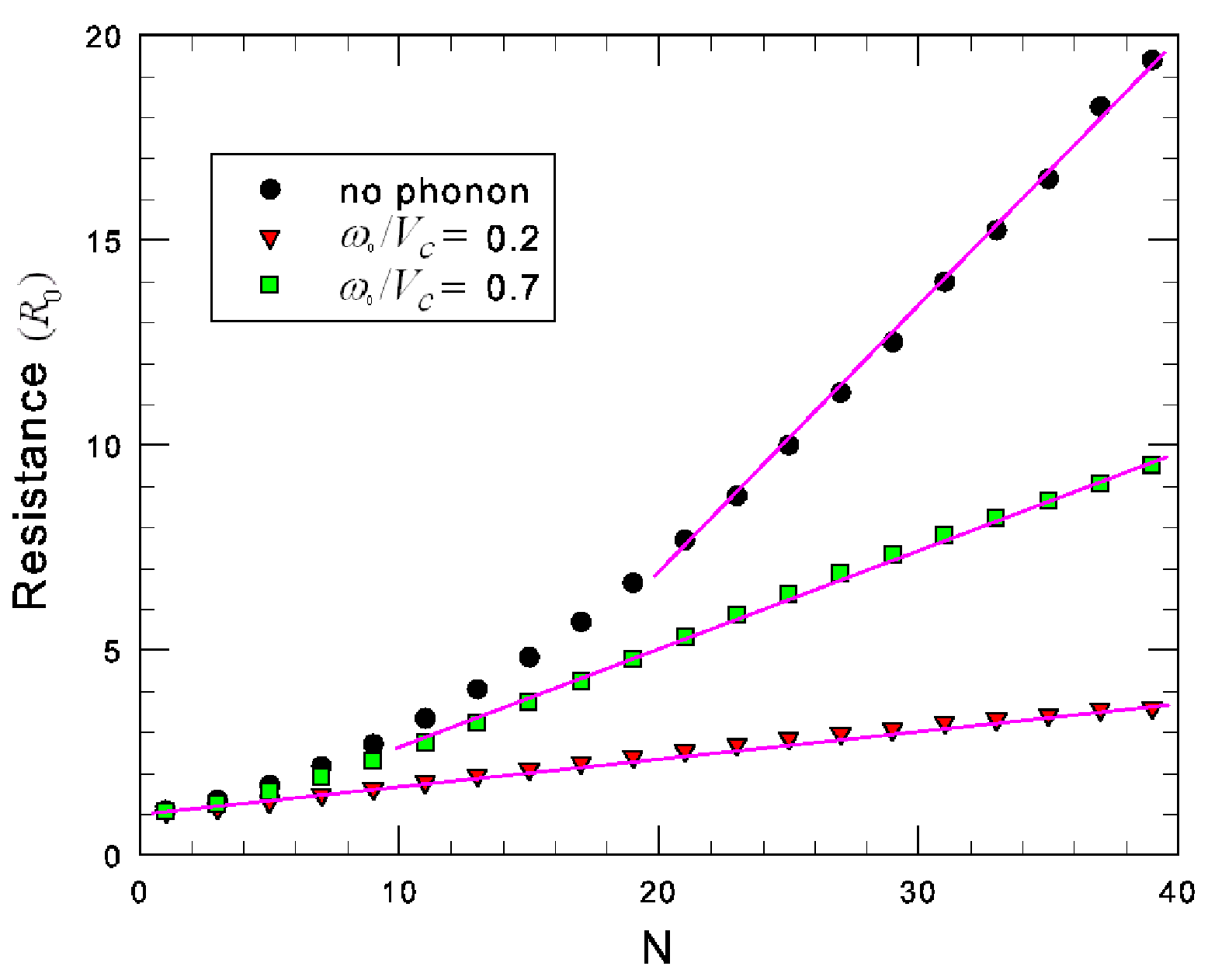}{(Color online) Resistance as a function of length  at $V_a=0.1$V. Circles corresponds to $\gamma=0$ , and triangles (squares) corresponds to $\gamma=0.1$ with $\omega_0= 0.2 $ (0.7). Solid lines show the linear relation between the resistance and  the chain length. }{hopping_Resistance_ga01_hw02_va01_2units}


\figuremacroW{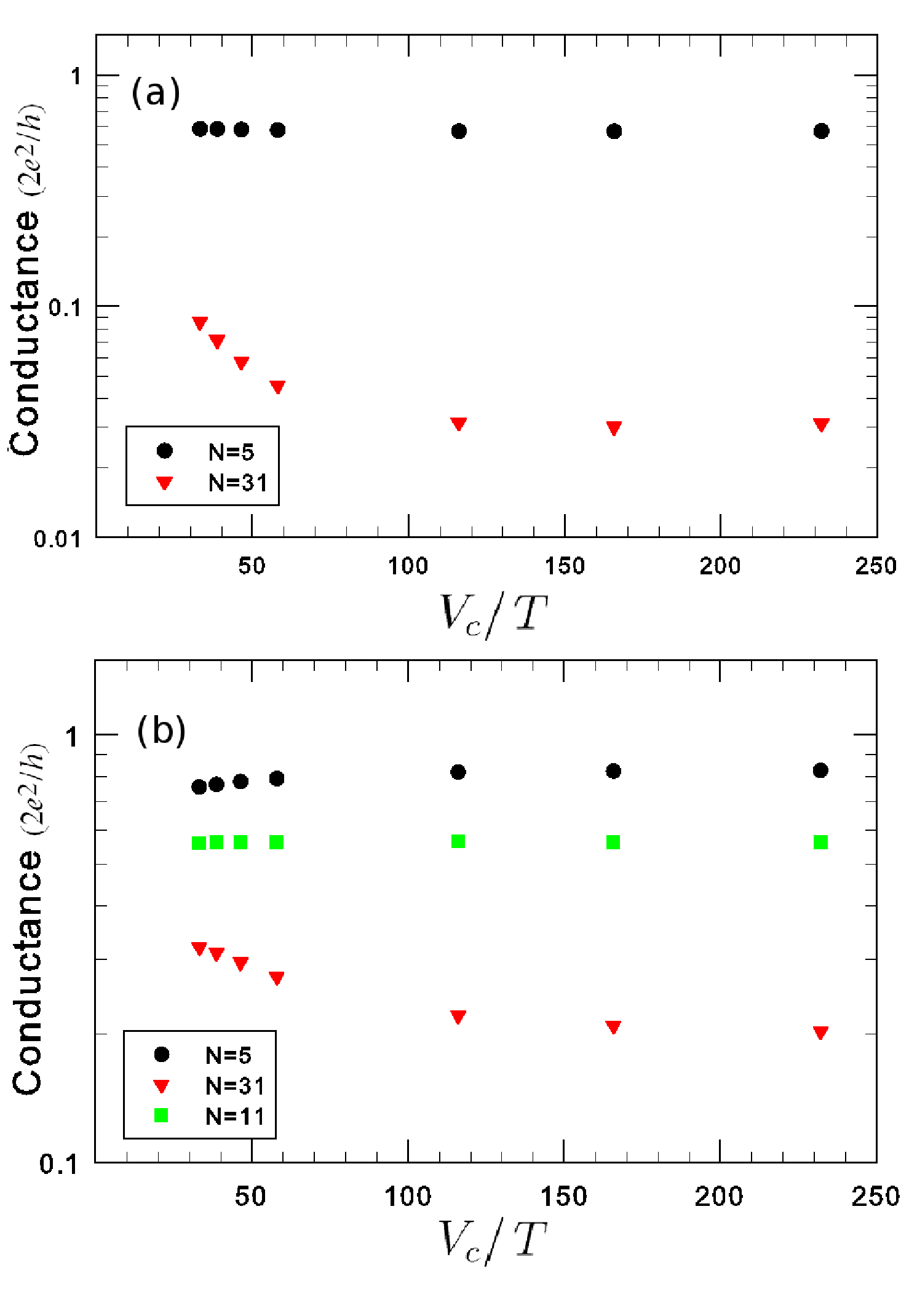}{(Color online) Conductance as a function of the inverse temperature for different lengths: $N=5$(circle),  11 (square), and 31 (triangle)  at $V_a=0.1$.  $\omega_0= 0.2$ and  (a) $\gamma=0$ and (b) $\gamma=0.1$ .}{temperature_dependence_2units}

\figuremacroW{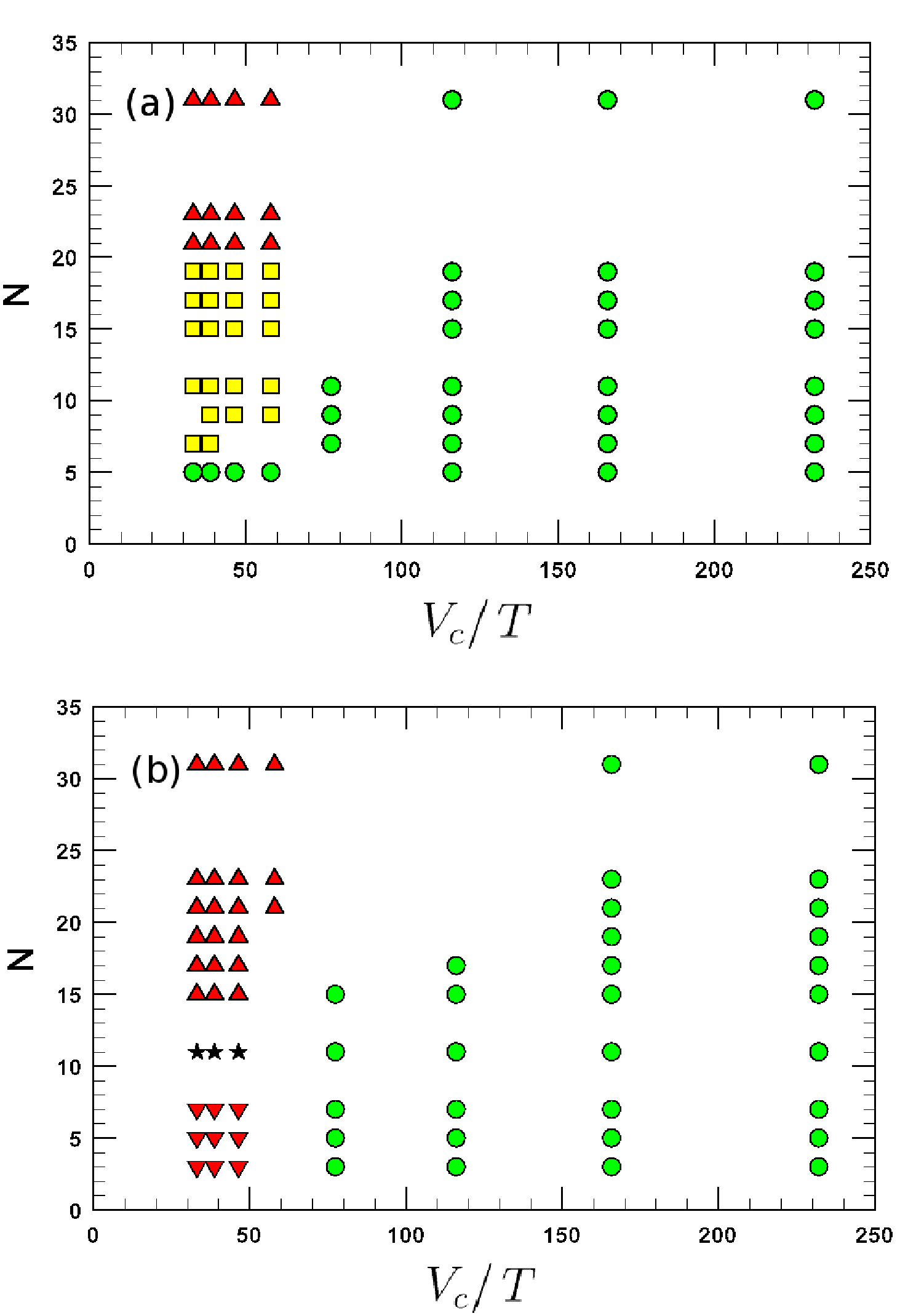}{(Color online) Charge transport mechanism  for different length and inverse temperature at $V_a=0.1$ for (a)  $\gamma=0$. Circles indicate tunneling transport, triangles correspond to thermally activated transport, and squares represent the crossover regime with the mixture of tunneling and thermally activated transport characteristics (see text), and  (b) for $\gamma=0.1$ and $\omega_0=0.2$.  The up and down triangles  indicate thermally assisted and suppressed  conduction, respectively, and stars represent a transition between these two behaviors.}{PHASE_transition}

 Figure~\ref{PHASE_transition} summarizes the conduction mechanism determined as described above for different temperatures and chain lengths. It is clear that at low temperature tunneling dominates. At high temperature, a change in the transport mechanism from tunneling to thermally activated transport occurs as the length increases.
Experiments show that the electron transport mechanism changes abruptly at a particular length.\cite{tuneling_to_hopping_2008_scinece_OPI,Frisbie_tTOh_2010_JACS_ONI,Lu_2009_OPE,Frisbie_tTOh_2010_JACS_OPT} 
However, our calculation [Fig.~\ref{PHASE_transition} (a)] shows that a crossover region appears between $N$=7 and $N$=19,
where the characteristics of tunneling and thermally activated transport coexist.\cite{REVIEW1}
Interestingly, for long chain molecules, there exists another transition from tunneling to thermally activated transport as the temperature increases. For short molecules, however, no such transition is observed and tunneling is still the main transport mechanism at high temperature.

The electron-phonon interaction can significantly change the transport characteristics.  Figure~\ref{tunneling_Conductance_ga01_hw02_va01_2units} shows the conductance versus length for 
$\gamma=0.1$, with phonon frequencies $\omega_0=0.2$ (triangles), 0.7(squares).  The inverse decay length $\beta$ decreases in the presence of the electron-phonon interaction for short chains, suggesting the coherent electron tunneling becomes weaker as the interaction with phonons is turned on. Increasing $\omega_0$ decreases the probability of the phonon processes, and the behavior of the conductance  reduces back to the non-interacting case ($\gamma=0$ or $\omega_0\rightarrow \infty$). 
In addition, for $\omega_0=0.2$, the resistance is proportional to the length even in short chains (Fig.~\ref{hopping_Resistance_ga01_hw02_va01_2units}) indicating thermally activated transport for all molecular sizes.
Temperature dependence of the conductance [Fig.~\ref{temperature_dependence_2units} (b)] shows an interesting crossover from a thermally assisted transport in the long molecule ($N=31$) to  a thermally suppressed transport in the short molecule ($N=5$), and the transition occurs roughly at $N=11$. 
This crossover of the temperature dependent conduction as length increases replaces the change of the transport mechanism in the phonon free case [Fig.~\ref{PHASE_transition} (b)]. 

To better understand the temperature dependence of the 
 conductance at bias $V_a$, formally defined as 
\begin{align}
 g(V_a) &=\lim_{\epsilon^+\rightarrow 0} \frac{I(V_a+\epsilon^+)-I(V_a-\epsilon^+)}{2\epsilon^+}\;,
\end{align}
where $\epsilon^+$ is an infinitesimal positive number, we perform further analysis as follows.
When the electron-phonon coupling is weak, the first term in the current definition (Eq.~(\ref{eq:current})) dominates and the density of state is  almost the same for the two bias voltages separated by the infinitesimal  $\epsilon^+$, thus
\begin{align}
 g(V_a) & \approx \lim_{\epsilon^+\rightarrow 0}\frac{e}{4h\epsilon^+} \int d\omega \;{\rm Tr}\{ \tilde{\mathbf{F}}(V_a)  \mathbf{D}(\omega)\} \;,\nonumber\\
   \tilde{\mathbf{F}}(V_a,T,\omega) &= \tilde{\mathbf{f}}(V_a+\epsilon^+,T,\omega)-\tilde{\mathbf{f}}(V_a-\epsilon^+,T,\omega)\;,\nonumber\\
   \tilde{\mathbf{f}}(V_a,T,\omega) &=f_{L}(\omega,\mu_L)\mathbf{\Gamma}_{L}-f_{R}(\omega,\mu_R)\mathbf{\Gamma}_{R}\;,\nonumber\\
   \mathbf{D}(\omega)&=i[\mathbf{G}^{r}(\omega)-\mathbf{G}^{a}(\omega)]\;,
\end{align}
where ${\rm Tr} \mathbf{D}(\omega)$ gives  the total density of states of the molecular chain, including the effects from electron-phonon coupling; $\tilde{\mathbf{f}}(V_a,T,\omega)$ is the net electron distribution from two leads at temperature $T$ and bias $V_a$. The derivative of the  electron distribution at  $V_a$ is given by $\lim_{\epsilon^+\rightarrow 0} \tilde{\mathbf{F}}(V_a,T,\omega)/2\epsilon^+$.  We define a parameter $\Upsilon$ as 
\begin{align}
\label{y}
\Upsilon(V_a,T_1,T_2) &= \int d\omega {\rm Tr}\{\Delta\tilde{\mathbf{F}}(V_a,T_1,T_2,\omega)\mathbf{D}(\omega)
 \}\nonumber\\
 &=\int d\omega \Delta\tilde{F}_{11}(\omega) D_{11}(\omega)+\Delta\tilde{F}_{NN}(\omega) D_{NN}(\omega)
\end{align}
where the difference between distributions at two different temperatures is $\Delta\tilde{\mathbf{F}}(V_a,T_1,T_2,\omega)=\tilde{\mathbf{F}}(V_a,T_1,\omega)-\tilde{\mathbf{F}}(V_a,T_2,\omega) $, and has only two  non-zero matrix elements  $\Delta\tilde{F}_{11}$, and $\Delta\tilde{F}_{NN}$ due to the coupling to the left lead $\Gamma^L_{11} $ and to the right lead $\Gamma^R_{NN}$ respectively.

The physics of a thermally activated or suppressed transport  is thus dictated by  $\Upsilon$.
In the case $T_1> T_2$,  $\Upsilon(V_a,T_1,T_2) $ is positive for  thermally activated transport , and negative for thermally suppressed transport. 
 Since the density of states has small temperature dependence, $\Delta \tilde{F}(\omega)$ would play the major role on the temperature behavior. Assuming the density of states at sites $1$ and $N$  are approximately equal due to symmetry, we can rewrite Eq.~(\ref{y}) as 
  \begin{align*}
\Upsilon(V_a,T_1,T_2)  &\approx \int d\omega \left[\Delta\tilde{F}_{11} (\omega)+\Delta\tilde{F}_{NN}(\omega)\right] D_{11}(\omega)\nonumber\\
&=\int d\omega \Delta\tilde{F}(\omega)D_{11}(\omega),
 \end{align*}
where $\tilde{F}=\tilde{F}_{11}+\tilde{F}_{NN}$. 
 Inset of Fig.~\ref{Fig11} shows the  $\Delta\tilde{F}(V_a,T_1=300K,T_2=70K,\omega) $ as a function of energy, We notice $\Delta\tilde{F}$ is negative near $\omega=-V_a$, and becomes positive away from this region.  Figure~\ref{Fig11} shows the integrand of Eq.~(\ref{y}).  For $N=5$, there is more negative area than the positive one, which gives  $\Upsilon<0$, and the transport is thermally suppressed. On the other hand, for  the $N=31$ chain, there is more positive area than the negative one,  resulting in a thermally activated transport.  
In summary, the temperature dependence of the conductance comes principally from $\Delta\tilde{F}(\omega)$, and  the characteristics of the molecules and the electron-phonon effects enter through the density of states $D(\omega)$. For $N=5$, the weak electron-phonon interaction results in a shift of the density of states toward $\omega=-V_a$. This  picks up more negative contribution in $\Delta\tilde{F}$, resulting in a thermally suppressed transport.

\figuremacroW{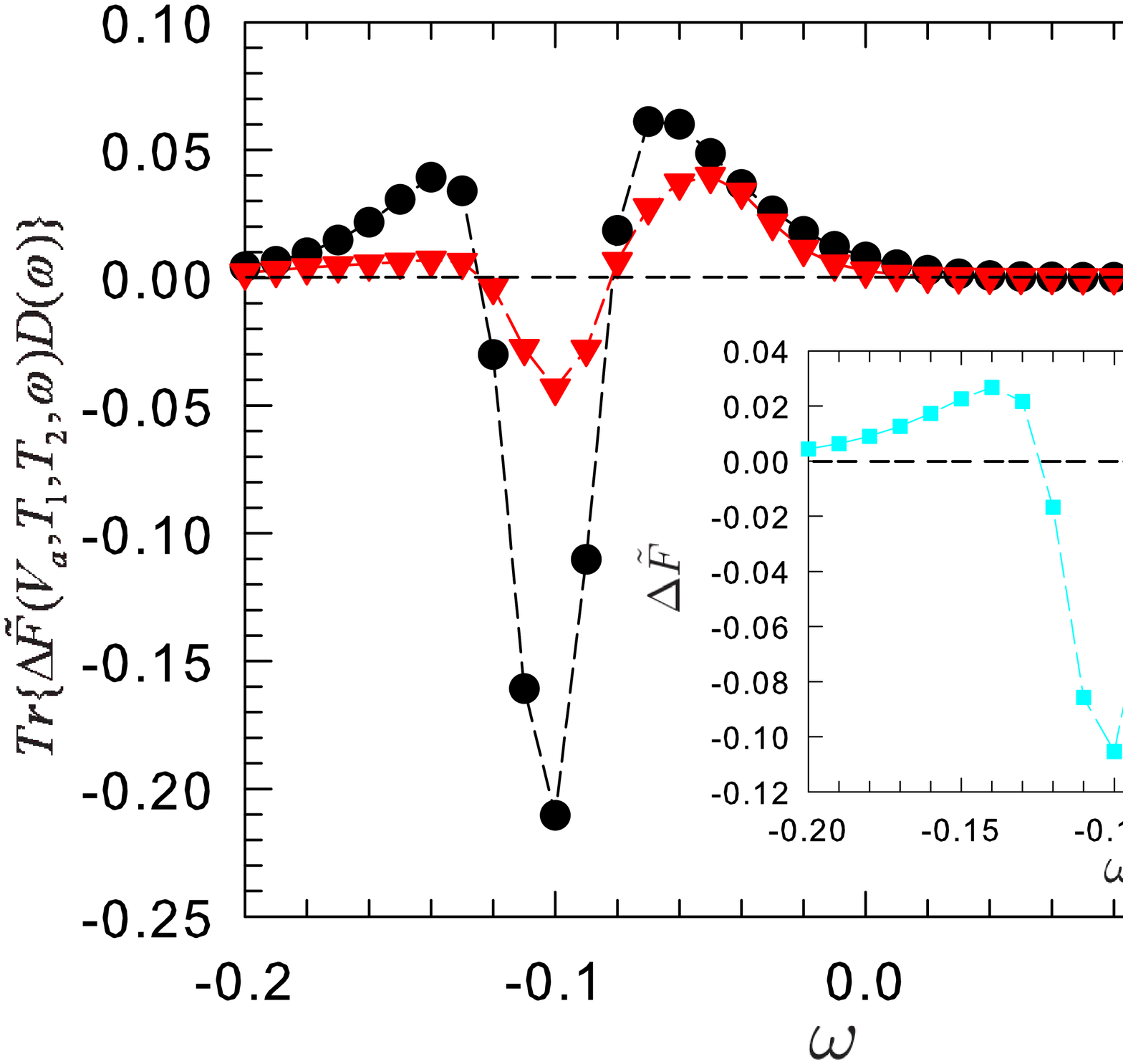}{(Color online) The integrand in Eq.~(\ref{y}) as a function of energy $\omega$ for $N=5$ (circle) and $N=31$ (triangle) at $V_a=0.1$, and $T_1$=300K, $T_2$=70K. Inset: Temperature difference of the effective electron distributions of the leads $\Delta\tilde{F}(\omega)$. See text for details. }{Fig11}
%

\section{CONCLUSION}\label{conclusion}
In conclusion, using non-equilibrium Green's function method, we study the length and temperature dependences of electron transport in molecular junctions. We model the molecular wire using a simple tight-binding model, with each site coupled locally to a phonon bath. We treat the electron-phonon interaction perturbatively up to second order in the electron-phonon coupling. 

We find different behaviors of conductance in short chains.   Larger conductance is observed in odd-numbered chains, and a resonance is present when the electron-phonon coupling increases. On the other hand, for even-numbered chains, low conductance is observed with a monotonic increase as the electron-phonon coupling increases.  In the absence of the electron-phonon interaction,  we observe a length dependent crossover of the conduction mechanism from tunneling to thermally activated transport at high temperature. We find that at high temperatures, tunneling transport dominates in short chains, while thermally activated transport dominates in long chains. For a long molecule of fixed length, there exists also a crossover from tunneling at low temperature to thermally activated transport at high temperature. In the presence of the electron-phonon interaction,  the coherent tunneling is interrupted by the phonon and the length crossover disappears. The 
thermally activated transport becomes the main transport mechanism. At high temperature a crossover of temperature dependent conduction from the  thermally suppressed  to assisted transport as the length increases, in place of the crossover in the transport mechanism in the phonon free case.

Without resorting to the \textit{ab initio} machinery, we are able to  extract essential physics in these molecular junctions using a simple model, and the results qualitatively agree with a wide range of experiments on molecular wires. The approach in this paper can be generalized to study molecules with site-dependent coupling strengths and phonon energies. This may be important to understand the mechanical roles of the function groups.  In addition, the results can be extended to the spin dependent case to study the properties of the  magnetic tunneling
junctions. Further improvement can also be made by self-consistently calculating the charge density at each site. These questions will be addressed in a future study. 

\begin{acknowledgments}
We thank Serguei Brazovski and Natasha Kirova for useful discussions.
We are grateful to the National Center for High-performance Computing for computer time and facilities.
This work was supported by the National Science Council in Taiwan through Grant No. NSC-98-2112-M-390-001-MY3, NSC-97-2120-M-006-010- (SJS), NSC-97-2628-M-002 -011-MY3, 99-2120-M-002-005- (YLL and YJK), and by NTU
Grant numbers 99R0066-65 and 99R0066-68 (YJK). The hospitality of the National Center for Theoretical Sciences, Taiwan,
where the work was initiated, is greatly acknowledged.
\end{acknowledgments}
\appendix
\section{}
\label{APPsolvingschr}
In this appendix, we provide an alternative method to study the transport properties of molecular chains by solving coupled Sch\"{o}dinger equations in the tight-binding limit for the on-site wave functions. For illustrative purposes, we consider the molecular transport without the electron-phonon interaction, and the full Hamiltonian can be written as,
\begin{align}
H=H_l+H_c+H_{el},
\end{align}  
where the Hamiltonian of two leads $H_l$ , the molecular electronic system $H_{el}$ , and the coupling between them $H_c$ are described by Eq.~(\ref{HL}) to (\ref{He}).
We solve the \textit{site-dependent} Schr\"odinger equations $E\phi_i=\sum_j H_{ij}\phi_j$ with the tight-binding model to obtain the transmission and reflection amplitudes. In the leads,  the dispersion relation $E-\bar\epsilon_\alpha= 2\eta_\alpha \cos(k_\alpha a_\alpha)$ for  $\alpha=L, R$ is employed, where $\bar\epsilon_\alpha$ includes both the on-site energy and the chemical potential in $\alpha$ lead, $a_\alpha$ is lattice constant, and $k_l$ ($k_r$) is incoming (outgoing) wave vector. We use plane waves as the wave functions in leads,
\begin{align}
	\phi_i^L &=a_1 e^{ik_lx_i^l}+a_2e^{-ik_lx_i^l}\;,\\
	\phi_i^R &=a_3 e^{ik_rx_i^r}\;,
\end{align}
where in the left lead the total wave function $\phi_i^L$ consists of  two components of the incoming and reflected waves, and  the transmitted wave in the right lead. The transition and reflection probabilities are defined as 
\begin{align}
	T &=c\left|\frac{a_3}{a_1}\right|^2\;,\\
	R &=c\left|\frac{a_2}{a_1}\right|^2\;,
\end{align}
where $c$ is a normalization constant given by  $|T|^2+|R|^2=1$. The current across the molecular junction is given by
\begin{align}
	I=\int dE\; T(E)[f_L(E,\mu_L)-f_R(E,\mu_R)]\;.
\end{align}

After solving the Schr\"odinger equations, the transition amplitude $a_3/a_1$ and the reflection amplitude $a_2/a_1$ are obtained by, 
\begin{widetext}
\begin{align}
	\frac{a_3}{a_1} &= \frac{2\eta_L V_LV_R \cos(k_la_l)e^{ik_lx_1^l}e^{-ik_rx_1^r}}{(E-\epsilon_R-\eta_Re^{ik_ra_r})[ V_L^2-(E-\epsilon_L-\eta_Le^{ik_la_l})(E-\epsilon_1-tD_2)]}\;\left(\frac{\phi_N^c}{\phi_1^c}\right)\;,\\
	\frac{a_2}{a_1} &=\frac{(E-\epsilon_L-\eta_Le^{-ik_la_l})(E-\epsilon_1-tD_2)-V_L^2}{V_L^2-(E-\epsilon_L-\eta_Le^{ik_la_l})(E-\epsilon_1-tD_2)}\;e^{2ik_lx_1^l}\;.
\end{align}
\end{widetext}
The ratio of wave functions at  the $N$th and the first site  is obtained by solving  the site-dependent Schr\"odinger equations of the molecular chain, which gives
\begin{align}
	\frac{\phi_N^c}{\phi_1^c}=\prod_{n=2}^N D_n\;,
\end{align}
where the Green's function on the $n$th site is given by 
\begin{align}
	D_n=\frac{t}{E-\epsilon_n}\;,
\end{align}
which propagates the wave function from the $n-1$th to the $n$th site  as $\phi_{n}^c =D_{n} \phi_{n-1}^c$.
Moreover, the molecular chain is  influenced by leads through the weak coupling $V_\alpha$ , and these effects are included in the self-energies.   We replace the on-site energies at the ends of the chain ($n=1, N$) with effective on-site energies, which  include the  self-energies from the leads. For example, the Schr\"odinger equation at site $N$ is given by, 
\begin{align}
	E\phi_N^c=t\phi_{N-1}^c+\epsilon_N\phi_N^c+V_R\phi^L_1\;,
\end{align}
and the Green's function $D_N=t/(E-\bar\epsilon_N)$ is obtained with the effective on-site energy $\bar\epsilon_N=\epsilon_N+V_R\phi^L_1/\phi_N^c$ , where the  ratio $\phi^L_1/\phi_N^c$ is obtained by solving the Schr\"odinger equation of the right lead,
\begin{align}
	\frac{\phi^L_1}{\phi_N^c}=\frac{V_R}{E-\epsilon_R-\eta_R e^{ik_ra_r}}\;.
\end{align}
On the other hand, the effective on-site energy $\bar\epsilon_N$ can also be described by the self-energy from the right lead. $\bar\epsilon_N=\epsilon_N+\Sigma_R$ , and $\Sigma_R$ can be obtained from Eq.~(\ref{leadselfenergies}).

\bibliographystyle{apsrev4-1}
\bibliography{ref_phonon}

\begin{thebibliography}{42}
\bibitem [{hop()}]{hoping.activation.transport}%
  \BibitemOpen
  \href@noop {} {}\bibinfo {note} {The two terms "thermally activated
  conduction" and "hopping transport" are used interchangeably in the
  literature. However, for hopping transport to occur, localized states are
  needed.}\BibitemShut {Stop}%
\end{thebibliography}
\end{document}